# Ten-Year Cross-Disciplinary Comparison of the Growth of Open Access and How it Increases Research Citation Impact

Chawki Hajjem < *hajjem.chawki@uqam.ca* >
Stevan Harnad < *harnad@uqam.ca* >
Yves Gingras < *gingras.yves@uqam.ca* >
Institute of Cognitive Sciences
Université du Québec à Montréal
Montréal, Québec, Canada H3C 3P8
http://www.er.uqam.ca/nobel/cogsci2/isc/

**Abstract**
*In 2001, Lawrence found that articles in computer science that were openly accessible (OA) on the Web were cited substantially more than those that were not. We have since replicated this effect in physics. To further test its cross-disciplinary generality, we used 1,307,038 articles published across 12 years (1992-2003) in 10 disciplines (Biology, Psychology, Sociology, Health, Political Science, Economics, Education, Law, Business, Management). We designed a robot that trawls the Web for full-texts using reference metadata (author, title, journal, etc.) and citation data from the Institute for Scientific Information (ISI) database. A preliminary signal-detection analysis of the robot's accuracy yielded a signal detectability d'=2.45 and bias £] = 0.52. The overall percentage of OA (relative to total OA + NOA) articles varies from 5%-16% (depending on discipline, year and country) and is slowly climbing annually (correlation r=.76, sample size N=12, probability p < 0.005). Comparing OA and NOA articles in the same journal/year, OA articles have consistently more citations, the advantage varying from 36%-172% by discipline and year. Comparing articles within six citation ranges (0, 1, 2-3, 4-7, 8-15, 16+ citations), the annual percentage of OA articles is growing significantly faster than NOA within every citation range (r > .90, N=12, p < .0005) and the effect is greater with the more highly cited articles (r = .98, N=6, p < .005). Causality cannot be determined from these data, but our prior finding of a similar pattern in physics, where percent OA is much higher (and even approaches 100% in some subfields), makes it unlikely that the OA citation advantage is merely or mostly a self-selection bias (for making only one's better articles OA). Further research will analyze the effect's timing, causal components and relation to other variables, such as, download counts, journal citation averages, article quality, co-citation measures, hub/authority ranks, growth rate, longevity, and other new impact measures generated by the growing OA database.*

# 1 Introduction

With the advent of the Internet and the Web, more and more researchers are making their research openly accessible (OA) by self-archiving it online [8, 18] to increase its visibility, usage and citation impact [5, 6, 16]. In 2001, Lawrence reported that OA articles in computer science are cited more. We have since replicated this OA citation advantage based on a single large central OA archive in physics [10, 11] and have begun testing it more widely [7]. We here report the generality of this effect across biological and social sciences, using a robot that trawls the Web for full-texts based on reference and citation data from the Institute for Scientific Information (ISI) database.

# 2 Method

Using the reference metadata for 1,307,038 articles published in peer-reviewed journals covered by the CDROM version of ISI's Science and Social Science Citation Indices (SCI and SSCI), our robot trawled the Web to estimate how many of the articles did (OA) or did not (NOA) have a full-text version freely accessible on the web. The 10 disciplines covered were: administration, economics, education, business, psychology, health, political science, sociology, biology, and law, for 12 years: 1992-2003.

The robot's search algorithm was the following: (1) Send request to ISI database for metadata of article (firstauthor name and article title). (2) Send request (name, title) to: Yahoo, Metacrawler, Vivissimo, Eo, AlltheWeb and Altavista. (3) Extract external (irrelevant) links. (4) Remove duplicate URLs. (5) Sort URLs to process PDF and PS files first (probable full-texts). (5) Convert files (PDF, PS, Latex, HTML, XML, RTF, and Word) to text. (6) Parse files to test for full-text of reference article (name/title in first 20% of text, references in last 20%). (7) If, in parsing HTML file, title found but not full text, extract and follow links in file further as references possibly leading to the full text (to depth of 3 levels). (8) Sort articles by discipline/journal/issue/year; calculate percent OA articles within each; then by discipline/journal; and finally for each discipline. (9) Sort articles by discipline/journal/issue/year, calculate citation ratio as (OA - NOA/NOA) within each, then by discipline/journal and finally for each discipline. (10) Exclude data for all journals that are 100% OA (OA journals) from both the article counts and the citation counts (as we are only doing within-journal comparisons for NOA journals); exclude data from all single issues that are 100% OA (to eliminate denominators).

# 3 Signal detection analysis of the robot's accuracy

To test the robot's accuracy, we performed a preliminary signal detection analysis [4]. From the 633,410 articles in Biology we took a sample of 100 articles the robot had called OA and 100 it had called NOA and handchecked them for correctness. This yielded four possibilities : *Hits* (correct positives: OA is called OA), *Correct rejections* (NOA is called NOA), *False alarms* (NOA is called OA) and *Misses* (OA is called NOA). In a sample of 100 articles tagged by the robot as OA and 100 tagged as NOA, the Robot had 6 Misses and 19 False Alarms according to a manual check of its accuracy.

Signal detectability (d') was found to be 2.45, indicating that the robot was fairly sensitive. The robot's bias £] = 0.52 indicates some tendency toward false alarms (overestimating OA). If £] = 1 the robot is neutral, favoring neither false alarms nor misses; £] > 1 favors misses and£] < 1 favors false alarms. As there are in fact about ten times as many NOA articles as OA articles, this means there is some overestimation of the percentage of OA articles and hence some underestimation of the size of any OA citation advantage we might find.

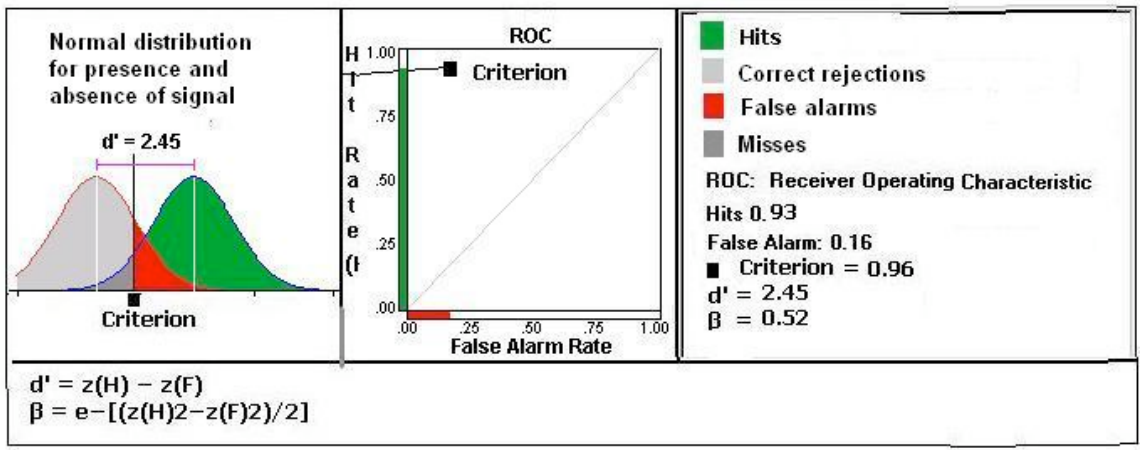

Figure 1: **Signal detection Analysis of robots Accuracy**. (Graph generated using the applet provided by Wise Project http://wise.cgu.edu/sdt/sdt.html )

| N=12 | r |
|---|---|
| OA Citation Advantage x Year | 0.25 NS |
| OA Citation Advantage x Total articles | 0.21 NS |
| OA Citation Advantage x %OA articles | -0.02 NS |
| Total articles x Year | 0.65 $p < 0.01$ |
| Total articles x %OA articles | 0.31 NS |
| %OA articles x Year | 0.76 $p < 0.005$ |

Table 1: **Correlation between Year and OA Growth.** Significant correlation between year and percent OA articles: %OA is growing annually. (Total articles is also growing yearly; no other correlations are significant.)

## 4 Results

Figure 2.a shows the 12-year average for the percentage of OA articles (dark bars) in each of our 10 reference disciplines, ordered by total number of articles (OA + NOA, with Biology on the high end and Law on the low end). Percent OA varies from 5%-16%. There is a clear and consistent OA citation advantage (OA-NOA/NOA calculated within each individual journal issue, then averaged across journals, but not counting issues that had 100% or 0% OA articles) across all the disciplines, varying from 36%-172% (white bars): OA articles have more citations. Figure 2.b shows that this OA citation advantage is present across all countries (based on 1st-author affiliation and ordered by total article output).

We now look more closely at the fine-structure of the OA citation advantage and OA growth across time. Figure 2.c shows pooled results across all the disciplines for total annual articles (OA + NOA, gray curve), percent OA (black bars, log scale) and percent OA citation advantage (white bars, log scale). Both total articles and annual percent OA are growing (slowly) from year to year (r=.65 and .76, respectively, Table 1; no other correlations are significant).

We next look at the time course of total percentage growth in OA (for all 10 disciplines) within specific citation ranges $OA_c$ (c= 0, 1, 2-3, 4-7, 8-15, 16+). Figure 3.a should be read backwards, 2003-1992, because citations grow with time, older articles accumulating more citations across the years. So it is perhaps not surprising that the percentage of OA articles among those articles with zero citations, $OA_0$ decreases with time (at first rapidly, from 2003 till about 1998, and then slowly leveling off). For articles with one or more citations, the corresponding effect is the opposite, $OA_c$ grows (backwards) with time (first rapidly from 2003 till about 1998, then likewise leveling off). But this is not a specific OA effect at all, for the inset shows the very same pattern is for NOA articles too. The specific OA effect only becomes apparent when we examine the corresponding ratio $OA_c/NOA_c$ within each citation range (Figure 3.b).

The OA effect only becomes apparent when we look at $OA_c/NOA_c$. This ratio is growing year by year

(a)

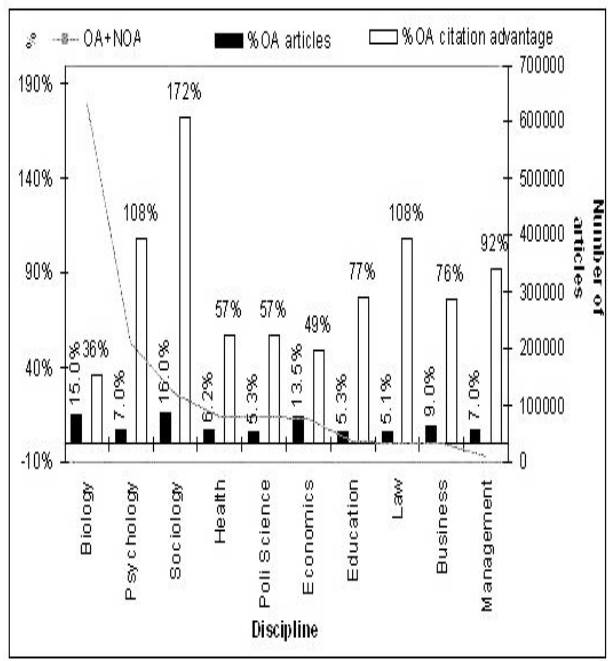

(b)

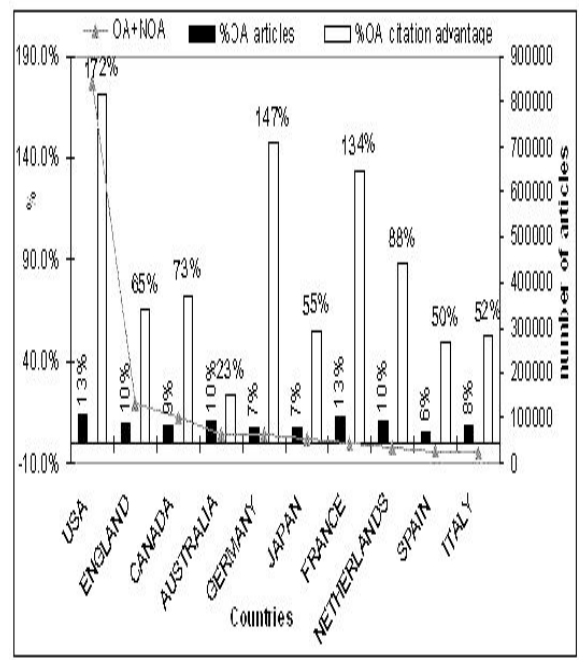

(c)

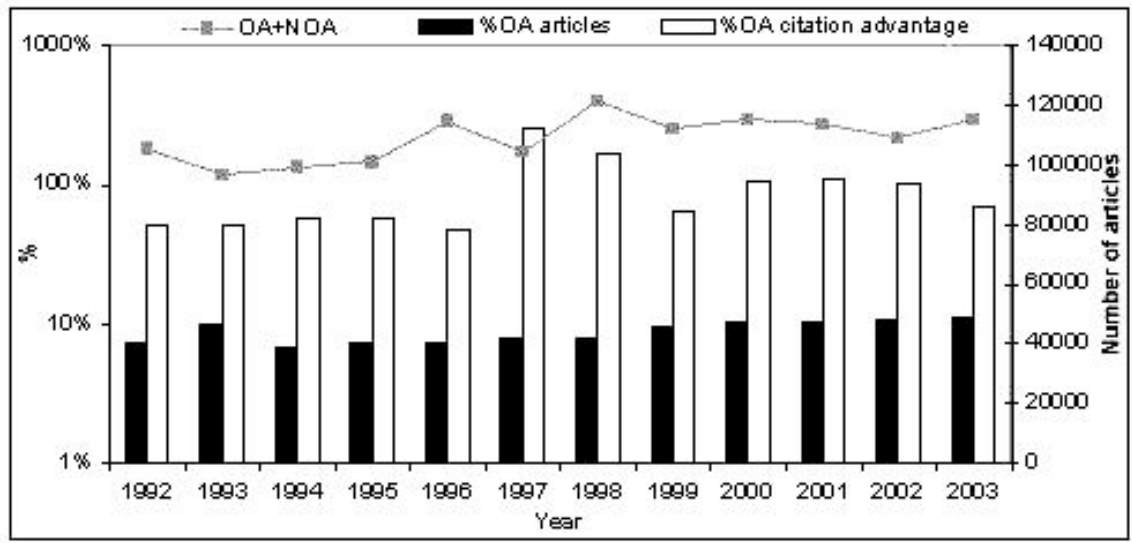

Figure 2: **(a): Open Access Citation Impact Advantage by Discipline.** Total articles (*OA+NOA*), gray curve; percentage OA: (*OA/(OA + NOA)*) articles, black bars; percentage OA citation advantage: ((*OA - NOA)/NOA*) citations, white bars, averaged across 1992-2003 and ranked by total articles. All disciplines show an OA citation advantage. **(b): Open Access Citation Impact Advantage by Country.** Total articles (gray curve), percent OA articles (black bars), and percent OA citation advantage (white bars); averaged across all disciplines and years 1992-2003; ranked by total articles. **(c): Open Access Citation Impact Advantage by Year.** Total articles (gray curve), percent OA articles (black bars), and percent OA citation advantage (white bars): 1992-2003, averaged across all disciplines. No yearly trend is apparent in the size of the OA citation advantage, but %OA is growing from year to year (see Table 1). Note that percent scale is logarithmic (to make the OA growth visible).

(Figure 3.b) which means that within each citation range, the percentage of articles that are OA is growing faster than the percentage of articles that are NOA (correlations are all positive and very high, Table 3). This growth differential also increases with the citation range, being lowest for uncited articles and highest for articles with over sixteen citations. This confirms the pattern reported for computer science articles by [15].

If we look at our total sample of 1,307,038 articles across all disciplines and years, we see that 793494 (61%) of them are uncited; of the remaining 513544 (39%), 155265 (12%) have 1 citation, declining to 53838 (4%) with 16+ citations (Figure 4, gray curve). 156845 (12%) of the total articles are OA. Of those, 85794 (55%) are uncited, and their numbers in each higher citation range fall off much the way the totals do (Figure 4, dark curve). However, if we again look at the ratios between the percentages among OA and NOA articles for each range, c, expressed as ($OA_c$-$NOA_c$)/$NOA_c$ (bars in Figure 4), we see that this ratio is positive for all nonzero citation ranges, beginning at 1 citation (16% OA advantage), peaking at about 4-7 (c. 22% OA advantage), and falling off again toward 16+ citations (10% OA advantage). This means that the proportion of articles within each citation range is greater among OA articles than among NOA articles except zero, the most populace category (61%), where it is NOA articles that have the -

12% NOA *dis*advantage.

In and of themselves, these correlations and temporal patterns cannot determine causality. It is a logical possibility that the cause of the OA advantage is merely a self-selection bias: that authors tend to self-archive their better papers (or better authors tend to self-archive their papers) and better papers are simply cited more.

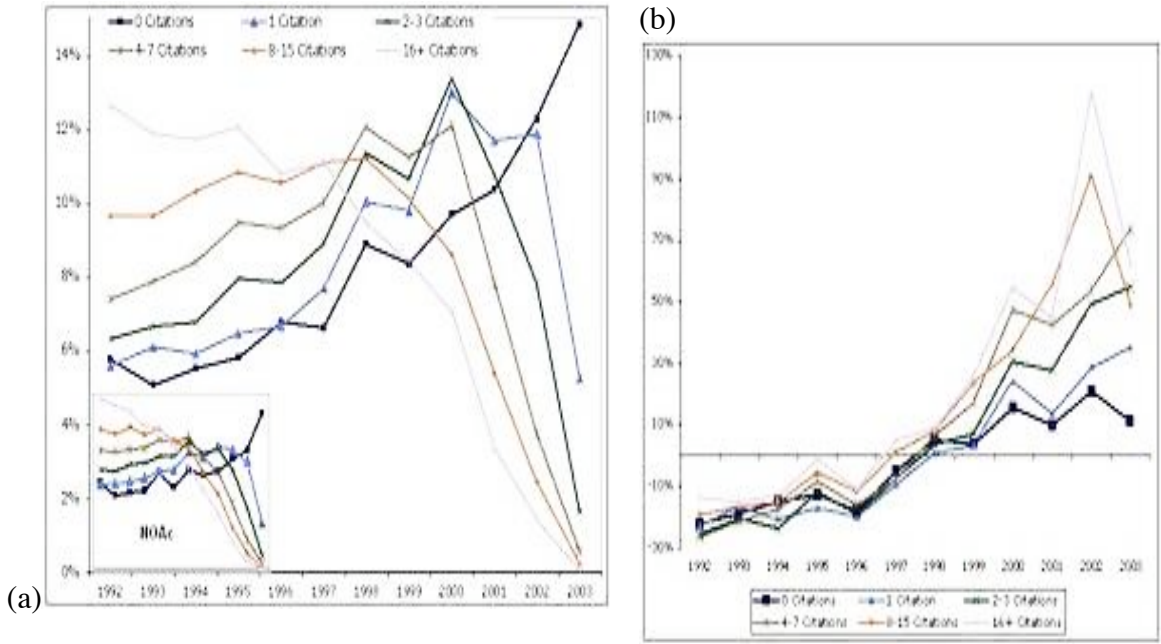

Figure 3: **(a): Yearly OA and NOA in each Citation Range.** The yearly percentage ($OA_c$) of the articles with c citations (c = 0, 1 2-3, 4-7, 8-15, 16+) that are OA (1992-2003). This graph (figure 3.a) should really be read backwards, as citations increase cumulatively as an article gets older (younger articles have fewer citations). Reading backwards, for articles with no citations (c=0), the percentage $OA_c$ decreases each year from 2003-1992, at first rapidly, then more slowly. For articles with one and more citations ($c > 0$), $OA_c$ first increases rapidly from 2003 till about 1998, then decreases slowly 1998-1992. Notice that the rank order becomes inverted around midway (c. 1998), the percentages increasing from c=0 to c=16+ for the oldest articles (1992) and the reverse for the youngest articles (2003). The pattern is almost identical for NOA articles too (see $NOA_c$ inset), so this is the relationship between citation ranges and time for all articles, not a specific OA effect. **(b): Yearly Growth of OA Relative to NOA in Each Citation Range.** The yearly ratio $OA_c/NOA_c$ between the percentage of articles with c citations (c = 0, 1 2-3, 4-7, 8-15, 16+) that are OA and NOA (all disciplines). This ratio is increasing with time (as well as with higher citation counts, c), showing that the effect first reported for computer science conference papers by Lawrence (2001) occurs for all disciplines.

This is unlikely to be the sole or even the primary cause of the OA advantage for three reasons,

two empirical and one commonsensical: (1) The first empirical reason is that if the OA advantage were solely a self-selection bias, it would have to shrink or disappear as the percentage of OA articles approaches 100%. Our sample's average percent OA content was low (around 9%), but prior studies in disciplines where the self-archiving rate is much higher – well over 50% in some areas of physics [10, 11] and near or at 100% in astronomy and astrophysics [12] – have found OA citation advantages that were of the same size as the ones found here. (2) The second empirical reason is that OA has also been shown to increase article downloads [1,**?**], and that increased downloads are in turn correlated with increased citations [2, 17, 19]. Causality is more directly evident there. (3) The commonsensical reason to assume that OA is causal is that access is a necessary (if not a sufficient) condition for usage and citation, and no researcher's institution can afford access to anywhere near all journals [http://www.arl.org/stats/arlstat/]; OA self-archiving supplements that access, increasing potential online accessibility to 100%.

| N=12 | r |
|---|---|
| O Citations $OA_c$ x Year | $0.94 p < 0.005$ |
| 1 Citations $OA_c$ x Year | $0.60 p < 0.025$ |
| 2 - 3 Citations $OA_c$ x Year | $0.10 p < 0.05$ |
| 4 - 7 Citations $OA_c$ x Year | $-0.36\ p < 0.05$ |
| 8 - 15 Citations $OA_c$ x Year | $-0.74\ p < 0.005$ |
| 16+ Citations $OA_c$ x Year | $-0.93\ p < 0.001$ |

Table 2: **Correlation between Year and Percent OA in Each Citation Range.** Significant correlations between year and the percentage of OA articles in each citation range, $OA_c$: Percent OA is growing annually (negative correlation) in the higher citation ranges and shrinking in the lower ones; but the correlation pattern is the same for NOA articles, hence this is not an OA effect. It just shows that citations increase with time.

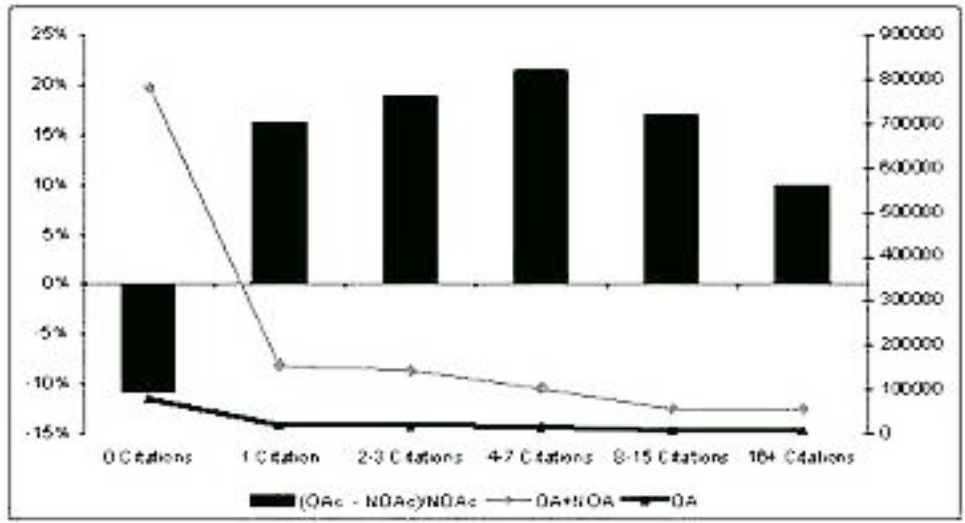

Figure 4: **OAc/NOAc Ratio in Each Citation Range (All years, All Disciplines).** Ratio of the percentage of articles with c citations (c = 0, 1 2-3, 4-7, 8-15, 16+) that are OA to the percentage that are NOA (across all disciplines and years), expressed as a difference from equality ($OA_c-NOA_c$)/$NOA_c$. This ratio increases as citation count (c) increases (r = .98, N=6, $p < .005$). The percentage of articles with 0 citations is relatively higher among NOA articles, but it becomes higher among OA articles with 1 citation and higher. This shows that the more cited an article, the more likely that it is OA. (The gray curve is the total number of articles (OA + NOA) in each citation range, and the dark curve is the number of OA articles scale for both curves is on right.)

## 5 Conclusion

Research is conducted (and funded and published) in order to be used, applied and built upon. It is for this reason that citation impact is rewarded by researchers' institutions and funders [3, 20]. It follows that whatever increases research access and impact increases benefits to research, researchers, their institutions and their funders. Our estimate of the current percentage of OA articles in the 10 disciplines tested is between 5% and 16% (mean 9%; median 7% ; SD 4.26) and that OA is associated with citation impact that is 36% to 172% higher (mean 83 %; median 77% ; SD 39.49). (Studies in further discipline [11] extend the range of %OA to 5%-15% and the range of the OA citation impact advantage to 25%-250%.) To extend this benefit to the remaining 85-95% of research, "publish or perish" needs to be extended, in the online age, to "publish and self-archive" so as to maximize research access and impact [21]. In addition to the direct impact benefits, as the OA database approaches 100%, many rich new measures of research usage and impact will become possible, including both citation and download counts, growth curves, and latencies; co-citation counts; hub/authority ranks, semantic indices [14] and many other online performance indicators. These will be usable not only for navigation and evaluation, but also for analyzing and predicting research directions and influences.

| N=12 | r |
|---|---|
| O Citations $OA_c/NOA_c$ x Year | 0.94 $p < 0.001$ |
| 1 Citations $OA_c/NOA_c$ x Year | 0.94 $p < 0.001$ |
| 2 - 3 Citations $OA_c/NOA_c$ x Year | 0.96 $p < 0.001$ |
| 4 - 7 Citations $OA_c/NOA_c$ x Year | 0.96 $p < 0.001$ |
| 8 - 15 Citations $OA_c/NOA_c$ x Year | 0.91 $p < 0.001$ |
| 16+ Citations $OA_c/NOA_c$ x Year | 0.87 $p < 0.001$ |

Table 3: **Correlation between Year and OAc/NOAc Growth Ratio in Each Citation Range.** Significant correlations between year (1992-2003) and the ratio $OA_c/NOA_c$ between the percentage of articles with c citations (c = 0, 1 2-3, 4-7, 8-15, 16+) that are OA and the percentage with c citations that are NOA (all disciplines). This ratio is growing annually in every citation range.